\documentclass[12pt,english]{article}
\usepackage[T1]{fontenc}
\usepackage[latin9]{inputenc}
\usepackage[a4paper]{geometry}
\usepackage[active]{srcltx}
\usepackage{textcomp}
\usepackage{amsmath}
\usepackage{amssymb}
\usepackage{setspace}
\PassOptionsToPackage{normalem}{ulem}
\usepackage{ulem}
\makeatletter

\newcommand{\lyxmathsym}[1]{\ifmmode\begingroup\def\b@ld{bold}
  \text{\ifx\math@version\b@ld\bfseries\fi#1}\endgroup\else#1\fi}

\newcommand{\lyxaddress}[1]{
\par {\raggedright #1
\vspace{1.4em}
\noindent\par}
}

\makeatother

\usepackage{babel}
\begin{document}

\title{Supersymmetrization of Quaternion Dirac Equation for Generalized
Fields of Dyons }

\author{A. S. Rawat\textsuperscript{(1)}, Seema Rawat\textsuperscript{(2) },
Tianjun Li$^{(3)}$ and O. P. S. Negi\textsuperscript{(3,4)}%
\thanks{Address for Correspondence during \textbf{Feb. 22-April 19, 2012}:
Institute of Theoretical Physics,Chinese Academy of Sciences, Zhong
Guan Cun East Street 55, P. O. Box 2735, Beijing 100190, P. R. China%
}}

\maketitle

\lyxaddress{\begin{center}
\textsuperscript{}1. Department of Physics, H. N. B. Garhwal University,
Pauri Campus, Pauri (Garhwal)-246001, Uttarakhand, India.\\
\textsuperscript{}2. Department of Physics, Zakir Husain College,
Delhi University, Jawaharlal Nehru Marg, New Delhi-110002, India.\\
3. Institute of Theoretical Physics,Chinese Academy of Sciences, Zhong
Guan Cun East Street 55, P. O. Box 2735, Beijing -100190, P. R. China.\\
4. Department of Physics, Kumaun University, S. S. J. Campus, Almora-
263601, Uttarakhand, India\\
email: 1. drarunsinghrawat@gmail.com; 2. rawatseema1@rediffmail.com;
3. tli@itp.ac.cn; 4. ops\_negi@yahoo.co.in
\par\end{center}}
\begin{abstract}
The quaternion Dirac equation in presence of generalized electromagnetic
field has been discussed in terms of two gauge potentials of dyons.
Accordingly, the supersymmetry has been established consistently and
thereafter the one, two and component Dirac Spinors of generalized
quaternion Dirac equation of dyons for various energy and spin values
are obtained for different cases in order to understand the duality
invariance between the electric and magnetic constituents of dyons. 


\end{abstract}

\section{Introduction:}

Symmetries are one of the most powerful tools in the theoretical physics.
Relativistic quantum mechanics is the theory of quantum mechanics
that is consistent with the Einstein\textquoteright{}s theory of relativity.
Dirac\cite{key-1} was the first who attempted in this field followed
by Feshback and Villars\cite{key-2}. Since relativistic quantum mechanics
in 3+1 space-time dimension becomes difficult because of different
dimensionality of time and space. Nevertheless, the use of quaternions
has become essential because quaternion algebra\cite{key-3} has certain
advantages. It provides 4-dimensional structure to relativistic quantum
mechanics and also provide consistent representation in terms compact
notations. Quaternions have direct link with Pauli spin matrices where
the spin \cite{key-4,key-5} plays an important role in order to make
connection between bosons and fermions. Pioneer work in the field
of relativistic quaternionic quantum mechanics was done by Adler\cite{key-4}
while Rotelli\cite{key-6} and Leo et al\cite{key-7,key-8} discussed
the quaternionic wave equation. Gürsey\cite{key-9} and Hestens\cite{key-10}
reformulated the Dirac equation from quaternion valued terms showing
that the algebraic equivalent of Dirac has been forced to break the
automorphism group of quaternions. Supersymmetric formulation of quaternionic
quantum mechanics \cite{key-4} has been discussed by Davies \cite{key-11}
into study supersymmetric quantum mechanics. More over, a lot of literature
has been cited \cite{key-12,key-13,key-14,key-15,key-16,key-17,key-18,key-19,key-20,key-21,key-22,key-23}
to describe the supersymmetry (SUSY) as the natural symmetry of spin
- particles. Nicolai \cite{key-24} has also introduced the SUSY for
spin system in statistical mechanics. Consequently, supersymmetric
method in quaternionic Dirac equation provides \cite{key-11} the
exact solutions of various problems. Keeping in view the advantages
of SUSY and the applications of quaternionic algebra, we \cite{key-25,key-26}
have also analyzed the supersymmertization of quaternion quantum mechanics
and quaternion Dirac equation for different masses. Extending our
results \cite{key-26} , in this paper, we have discussed the quaternion
Dirac equation in electromagnetic field where the partial derivative
has been replaced by the quaternion covariant derivative. The quaternion
Dirac equation in electromagnetic field consists of two gauge fields
subjected by two unitary gauge transformations in terms of two gauge
potentials. These two gauge potentials are identified as the gauge
potentials respectively associated with the simultaneous existence
of electric and magnetic charge ( particles named as dyons \cite{key-27,key-28}).
Accordingly, we have obtained the one and two components solutions
of generalized quaternion Dirac equation of dyons for its different
cases associated with its electric and magnetic constituents. Furthermore,
we have analyzed, the supersymmertization of generalized quaternion
Dirac equation of dyons for considering different cases of electric
and magnetic fields interacting with electric and magnetic charges
as the consequence of electromagnetic duality of dyons.

\section{Quaternion Preliminaries:}

The algebra $\mathbb{H}$ of quaternion is a four-dimensional algebra
over the field of real numbers $\mathbb{R}$ and a quaternion $\phi$
is expressed in terms of its four base elements as

\begin{align}
\phi=\phi_{\mu}e_{\mu}= & \phi_{0}+e_{1}\phi_{1}+e_{2}\phi_{2}+e_{3}\phi_{3}(\forall\mu=0,1,2,3)\label{eq:1}
\end{align}
where $\phi_{0}$,$\phi_{1}$,$\phi_{2}$,$\phi_{3}$ are the real
quartets of a quaternion and $e_{0},e_{1},e_{2},e_{3}$ are called
quaternion units and satisfies the following relations,

\begin{align}
e_{0}^{2} & =e_{0}=1,;\,\,\,\, e_{j}^{2}=-e_{0};\nonumber \\
e_{0}e_{i}=e_{i}e_{0} & =e_{i}(i=1,2,3);\nonumber \\
e_{i}e_{j} & =-\delta_{ij}+\varepsilon_{ijk}e_{k}(\forall\, i,j,k=1,2,3)\label{eq:2}
\end{align}
where $\delta_{ij}$ is the delta symbol and $\varepsilon_{ijk}$
is the Levi Civita three index symbol having value $(\varepsilon_{ijk}=+1)$
for cyclic permutation, $(\varepsilon_{ijk}=-1)$ for anti cyclic
permutation and $(\varepsilon_{ijk}=0)$ for any two repeated indices.
Addition and multiplication are defined by the usual distribution
law $(e_{j}e_{k})e_{l}=e_{j}(e_{k}e_{l})$ along with the multiplication
rules given by equation (\ref{eq:2}). $\mathbb{H}$ is an associative
but non commutative algebra. If $\phi_{0},\phi_{1},\phi_{2},\phi_{3}$
are taken as complex quantities, the quaternion is said to be a bi-
quaternion. Alternatively, a quaternion is defined as a two dimensional
algebra over the field of complex numbers $\mathbb{C}$. We thus have
$\phi=\upsilon+e_{2}\omega(\upsilon,\omega\in\mathbb{C})$ and $\upsilon=\phi_{0}+e_{1}\phi_{1}$
, $\omega=\phi_{2}-e_{1}\phi_{3}$ with the basic multiplication law
changes to $\upsilon e_{2}=-e_{2}\bar{\upsilon}$.The quaternion conjugate
$\overline{\phi}$ is defined as 

\begin{align}
\overline{\phi}=\phi_{\mu}\bar{e_{\mu}}= & \phi_{0}-e_{1}\phi_{1}-e_{2}\phi_{2}-e_{3}\phi_{3}.\label{eq:3}
\end{align}
In practice $\phi$ is often represented as a $2\times2$ matrix $\phi=\phi_{0}-i\,\vec{\sigma}\cdot\vec{\phi}$
where $e_{0}=I,e_{j}=-i\,\sigma_{j}(j=1,2,3)$ and $\sigma_{j}$are
the usual Pauli spin matrices. Then $\overline{\phi}=\sigma_{2}\phi^{T}\sigma_{2}$
with $\phi^{T}$ is the transpose of $\phi$. The real part of the
quaternion $\phi_{0}$ is also defined as

\begin{align}
Re\,\phi & =\frac{1}{2}(\overline{\phi}+\phi)\label{eq:4}
\end{align}
where $Re$ denotes the real part and if $Re\,\phi=0$ then we have
$\phi=-\overline{\phi}$ and imaginary $\phi$ is known as pure quaternion
written as

\begin{align}
\phi= & e_{1}\phi_{1}+e_{2}\phi_{2}+e_{3}\phi_{3}.\label{eq:5}
\end{align}
The norm of a quaternion is expressed as $N(\phi)=\phi\overline{\phi}=\overline{\phi}\phi=\sum_{j=0}^{3}\phi_{j}^{2}$which
is non negative i.e.

\begin{align}
N(\phi)=\left|\phi\right|= & \phi_{0}^{2}+\phi_{1}^{2}+\phi_{2}^{2}+\phi_{3}^{2}=Det.(\phi)\geq0.\label{eq:6}
\end{align}
Since there exists the norm of a quaternion, we have a division i.e.
every $\phi$ has an inverse of a quaternion and is described as

\begin{align}
\phi^{-1}= & \frac{\overline{\phi}}{\left|\phi\right|}.\label{eq:7}
\end{align}
While the quaternion conjugation satisfies the following property

\begin{align}
\overline{\phi_{1}\phi_{2}}= & \overline{\phi_{2}}\,\overline{\phi_{1}}.\label{eq:8}
\end{align}
The norm of the quaternion (\ref{eq:1}) is positive definite and
enjoys the composition law

\begin{align}
N(\phi_{1}\phi_{2})= & N(\phi_{1})N(\phi_{2}).\label{eq:9}
\end{align}
Quaternion (\ref{eq:1}) is also written as $\phi=(\phi_{0},\vec{\phi})$
where $\vec{\phi}=e_{1}\phi_{1}+e_{2}\phi_{2}+e_{3}\phi_{3}$ is its
vector part and $\phi_{0}$ is its scalar part. So, the sum and product
of two quaternions are described as

\begin{align}
(\alpha_{0}\vec{,\,\alpha})+(\beta_{0}\vec{,\,\beta}) & =(\alpha_{0}+\beta_{0},\,\vec{\alpha}+\vec{\beta});\nonumber \\
(\alpha_{0}\vec{,\,\alpha})\cdot(\beta_{0}\vec{,\,\beta}) & =(\alpha_{0}\beta_{0}-\overrightarrow{\alpha}\cdot\overrightarrow{\beta}\,,\alpha_{0}\overrightarrow{\beta}+\beta_{0}\overrightarrow{\alpha}+\overrightarrow{\alpha}\times\overrightarrow{\beta}).\label{eq:10}
\end{align}
Quaternion elements are non-Abelian in nature and thus represent a
non commutative division ring.

\section{Quaternion Dirac Equation For Dyons: }

The free particle quaternion Dirac equation is described \cite{key-6}
as,

\begin{align}
(i\,\gamma^{\mu}\partial_{\mu}- & m)\Psi(x,t)=0\label{eq:11}
\end{align}
where $\Psi(x,t)=\left(\begin{array}{c}
\Psi_{a}(x,t)\\
\Psi_{b}(x,t)
\end{array}\right)$ is the two component spinor and

\begin{align}
\Psi_{a}(x,t)=\Psi_{0}+e_{1}\,\Psi_{1}; & \Psi_{b}(x,t)=\Psi_{2}-e_{1}\,\Psi_{3}\,\label{eq:12}
\end{align}
are the components of spinor quaternion $\Psi=\Psi_{0}+e_{1}\,\Psi_{1}+e_{2}\,\Psi_{2}+e_{3}\,\Psi_{3}$
and Dirac $\gamma$ matrices are also expressed in terms of quaternion
units i.e.

\begin{align}
\gamma_{0} & =\left[\begin{array}{cc}
1 & 0\\
0 & -1
\end{array}\right];\,\,\,\,\,\,\,\,\,\,\,\,\gamma_{j}=ie_{j}\left[\begin{array}{cc}
0 & 1\\
-1 & 0
\end{array}\right](\forall j=1,2,3).\label{eq:13}
\end{align}
So, the set of pure quaternion field (\ref{eq:1}) remains invariant
under the transformations 

\begin{align}
\phi & \rightarrow\phi'=U\phi\overline{U},\,\,\,\,\,\,\,\,\,\, U\in Q,\,\,\,\, U\overline{U}=1.\label{eq:14}
\end{align}
Similarly, the quaternion conjugate $\overline{\phi}$ transforms
as

\begin{align}
\overline{\phi'} & =\overline{U\phi\overline{U}}=U\,\overline{\phi}\overline{U\,}=-U\phi\overline{U}=-\phi'\label{eq:15}
\end{align}
Any $U\in Q$ has a decomposition like equation (\ref{eq:14}) which
gives rise to a set $\{U\in Q;\,\,\,\, U\overline{U}=1\}\sim SP(1)\sim SU(2).$
Though it has been emphasized earlier \cite{key-4} that the automorphic
transformation of $Q-$fields are local but one is free to select
them according to the representations. On the other hand, a $Q-$field
is subjected to more general $SO(4)$ transformations as 

\begin{align}
\phi & \rightarrow\phi'=U_{1}\phi\overline{U}_{2},\,\,\,\,\,\,\,\,\,\, U_{1},U_{2}\in Q,\,\,\,\, U_{1}\overline{U_{1}}=U_{2}\overline{U_{2}}=1.\label{eq:16}
\end{align}
So, the covariant derivative may then be described \cite{key-4} in
terms of two $Q-$ gauge fields i.e

\begin{align}
D_{\mu}\phi & =\partial_{\mu}\phi+A_{\text{\textmu}}\phi-\phi B_{\text{\textmu}}\label{eq:17}
\end{align}
which is subjected by two gauges $A_{\lyxmathsym{\textmu}}$ and $B_{\lyxmathsym{\textmu}}$
transforming like

\begin{align}
A_{\text{\textmu}}' & =U_{1}A_{\text{\textmu}}\overline{U_{1}}+(\partial_{\mu}U_{1})\overline{U_{1}};\nonumber \\
B_{\text{\textmu}}' & =U_{2}B_{\text{\textmu}}\overline{U_{2}}+(\partial_{\mu}U_{2})\overline{U_{2}};\label{eq:18}
\end{align}
where $A_{\lyxmathsym{\textmu}}$ and $B_{\lyxmathsym{\textmu}}$
may be identified as the four potentials associated, respectively,
with the electric and magnetic charges of dyons in terms of $U(1)\times U(1)$
gauge theory \cite{key-28}. Here, the gauge transformations are Abelian
and global. The quaternion covariant derivative given by equation
(\ref{eq:17}) thus supports the idea of two four potentials of dyons.
Accordingly, we way write the Dirac equation (\ref{eq:11}) for dyons
on replacing the partial derivative $\partial_{\mu}$ by covariant
derivative $D_{\mu}$as 

\begin{align}
(i\,\gamma^{\mu}D_{\mu}- & m)\Psi(x,t)=0\label{eq:19}
\end{align}
where the commutator is defined as 

\begin{align}
\left[D_{\mu},\, D_{\nu}\right]\Psi & =D_{\mu}(D_{\nu}\Psi)-D_{\nu}(D_{\mu}\Psi)=F_{\mu\nu}\Psi-\Psi\widetilde{F_{\mu\nu}}.\label{eq:20}
\end{align}
Here the gauge field strengths $F_{\mu\nu}$ and $\widetilde{F_{\mu\nu}}$
are described \cite{key-28} as the generalized anti-symmetric dual
invariant electromagnetic field tensors for dyons and are expressed
as 

\begin{align}
F_{\mu\nu}= & \partial_{\nu}A_{\mu}-\partial_{\mu}A_{\nu}-\frac{1}{2}\varepsilon_{\mu\nu\lambda\sigma}(\partial^{\lambda}B^{\sigma}-\partial^{\sigma}B^{\sigma});\nonumber \\
\widetilde{F_{\mu\nu}} & =\partial_{\nu}B_{\mu}-\partial_{\mu}B_{\nu}-\frac{1}{2}\varepsilon_{\mu\nu\lambda\sigma}(\partial^{\lambda}A^{\sigma}-\partial^{\sigma}A^{\sigma});\label{eq:21}
\end{align}
which leads to the following expressions \cite{key-28} for the generalized
electromagnetic fields of dyons i.e.

\begin{align}
\mathrm{\overrightarrow{\mathrm{E}}} & =-\frac{\partial\overrightarrow{A}}{\partial t}-\overrightarrow{\nabla}\phi-\overrightarrow{\nabla}\times\overrightarrow{B};\nonumber \\
\overrightarrow{\mathrm{B}} & =-\frac{\partial\overrightarrow{B}}{\partial t}-\overrightarrow{\nabla}\varphi+\overrightarrow{\nabla}\times\overrightarrow{A};\label{eq:22}
\end{align}
where $\left\{ A_{\lyxmathsym{\textmu}}\right\} =\left\{ \phi,\,-\vec{A}\right\} $
and $\left\{ B_{\lyxmathsym{\textmu}}\right\} =\left\{ \varphi,\,-\vec{B}\right\} $.
Generalized electromagnetic field tensors (\ref{eq:21}) of dyons
satisfy the following famous covariant form of Generalized Dirac-Maxwell's
(GDM) equations in presence of magnetic monopoles\cite{key-1} i.e.

\begin{align}
F_{\mu\nu,\nu}= & j_{\mu};\nonumber \\
\widetilde{F_{\mu\nu,\nu}}= & k_{\mu};\label{eq:23}
\end{align}
where $\left\{ j_{\mu}\right\} =\left\{ \rho,\,-\overrightarrow{j}\right\} =\mathbf{e\,\bar{\Psi\gamma_{\mu}\Psi}}$and
$\left\{ k_{\mu}\right\} =\left\{ \varrho,\,-\overrightarrow{k}\right\} =\mathbf{g\,\bar{\Psi\gamma_{\mu}\Psi}}$are
described \cite{key-28} as the four currents respectively associated
with the electric $\mathbf{e}$ and magnetic $\mathbf{g}$ charges
of dyons. We may now expend the four potentials (gauge potentials)
in terms of quaternion as

\begin{align}
A_{\text{\textmu}} & =A_{\text{\textmu}}^{^{0}}e_{0}+A_{\text{\textmu}}^{^{1}}e_{1}+A_{\text{\textmu}}^{^{2}}e_{2}+A_{\text{\textmu}}^{^{3}}e_{3};\nonumber \\
B_{\mu}= & B_{\text{\textmu}}^{^{0}}e_{0}+B_{\text{\textmu}}^{^{1}}e_{1}+B_{\text{\textmu}}^{^{2}}e_{2}+B_{\text{\textmu}}^{^{3}}e_{3}.\label{eq:24}
\end{align}
As such, the Abelian theory of dyons can now be restored by taking
the real part of the quaternion (\ref{eq:24}) $A_{\lyxmathsym{\textmu}}=\overline{A_{\lyxmathsym{\textmu}}}$
and $B_{\lyxmathsym{\textmu}}=\overline{B_{\lyxmathsym{\textmu}}}$
implying that $(A_{\lyxmathsym{\textmu}}^{^{0}})'=(A_{\lyxmathsym{\textmu}}^{^{0}})=A_{\mu}$
and $(B_{\lyxmathsym{\textmu}}^{^{0}})'=(B_{\lyxmathsym{\textmu}}^{^{0}})=B_{\mu}$.
However, if we consider the imaginary quaternion i.e. $A_{\lyxmathsym{\textmu}}=-\overline{A_{\lyxmathsym{\textmu}}}$
and $B_{\lyxmathsym{\textmu}}=-\overline{B_{\lyxmathsym{\textmu}}}$
we have the $SU(2)\times SU(2)$ gauge structure where $A_{\lyxmathsym{\textmu}}=A_{\lyxmathsym{\textmu}}^{^{a}}e_{a}=A_{\lyxmathsym{\textmu}}^{^{1}}e_{1}+A_{\lyxmathsym{\textmu}}^{^{2}}e_{2}+A_{\lyxmathsym{\textmu}}^{^{3}}e_{3}$
and $B_{\lyxmathsym{\textmu}}=B_{\lyxmathsym{\textmu}}^{^{a}}e_{a}=B_{\lyxmathsym{\textmu}}^{^{1}}e_{1}+B_{\lyxmathsym{\textmu}}^{^{2}}e_{2}+B_{\lyxmathsym{\textmu}}^{^{3}}e_{3}$.
Thus, with the implementation of condition $U_{1}\overline{U_{1}}=U_{2}\overline{U_{2}}=1$
there are only the six gauge fields $A_{\lyxmathsym{\textmu}}^{^{a}}$and
$B_{\lyxmathsym{\textmu}}^{^{a}}$ associated with the covariant derivative
of Dirac equation (\ref{eq:19}). The transformation equation (\ref{eq:16})
is continuous and isomorphic to $SO(4)$ i.e.

\begin{align}
\overline{\phi'}\phi' & =\overline{(U_{1}\phi\overline{U}_{2})}(U_{1}\phi\overline{U}_{2})=U_{2}\overline{\phi}\,\overline{U_{1}}\, U_{1}\phi\overline{U_{2}}=U_{2}\overline{\phi}\phi\overline{U_{2}}=\overline{\phi}\phi.\label{eq:25}
\end{align}
The resulting $Q-$ gauge theory has the correspondence $SO(4)\sim SO(3)\times SO(3)$
isomorphic to $SU(2)\times SU(2)$. Accordingly, the spinor transforms
as left and right component (electric or magnetic) spinors as 

\begin{align}
\Psi_{\mathbf{e}} & \mapsto(\Psi_{\mathbf{e}})'=U_{1}\Psi_{\mathbf{e}}\,\,\,\,\&\,\,\,\,\,\Psi_{\mathbf{g}}\mapsto(\Psi_{\mathbf{g}})'=U_{2}\Psi_{\mathbf{g}}.\label{eq:26}
\end{align}
The following split basis of quaternion units may also be considered
as

\begin{align}
u_{0} & =\frac{1}{2}(1-i\, e_{3});\,\,\,,\,\,\,\,\,\, u_{0}^{\star}=\frac{1}{2}(1+i\, e_{3});\nonumber \\
u_{1} & =\frac{1}{2}(e_{1}+i\, e_{2});\,\,\,,\,\,\,\,\,\, u_{1}^{\star}=\frac{1}{2}(e_{1}-i\, e_{2});\label{eq:27}
\end{align}
to constitute the $SU(2)$ doublets. As such, we may express the $Q-$classes
into five groups and can expand the theory with these choices. These
five irreducible representations of $SO(4)$ are realized as 

\begin{align}
1. & (U_{1},U_{2})\Rightarrow SO(4)\mapsto(2,2)\nonumber \\
2. & (U_{1},U_{1})\Rightarrow SU(2)\mapsto(3,1)\nonumber \\
3. & (U_{2},U_{2})\Rightarrow SU(2)\mapsto(1,3)\nonumber \\
4. & (U_{1},1)\Rightarrow Spinor\mapsto(2,1)\nonumber \\
5. & (U_{2},1)\Rightarrow Spinor\mapsto(1,2).\label{eq:28}
\end{align}
Accordingly, it is easier to develop a non-Abelian gauge theory of
dyons. It is to be mentioned that the occurrence of two gauge potentials
supports the idea of duality invariance \cite{key-29} among the electric
and magnetic parameters of dyons.

\section*{3 Supersymmetrization of Quaternion Dirac Equation for Dyons }

Quaternion Dirac equation (\ref{eq:11}) for dyons may now be written
as 

\begin{align}
i\gamma_{\mu}D_{\mu}\psi\left(x,t\right)= & m\psi\left(x,t\right)\label{eq:29}
\end{align}
where $\gamma$ matrices satisfy the properties 

\begin{align}
\gamma_{0}^{2}=+1;\,\,\, & \gamma_{l}^{2}=-1\,(\forall l=1,2,3)\nonumber \\
\gamma_{\mu}\gamma_{\nu}+\gamma_{\nu}\gamma_{\mu}= & -2g_{\mu\nu}\,(g_{\mu\nu}=-1,+1,+1,+1)\label{eq:30}
\end{align}
showing that $\gamma_{0}$ is Hermitian while $\gamma_{l}$ are anti-Hermitian
matrices. Accordingly, the matrix $\gamma_{5}$ may be expressed as

\begin{align}
\gamma_{5}=\gamma_{0}\gamma_{1}\gamma_{2}\gamma_{3}= & \left[\begin{array}{cc}
0 & -1\\
1 & 0
\end{array}\right]\label{eq:31}
\end{align}
 which satisfies the relations 

\begin{align}
\gamma_{0}\gamma_{5}+\gamma_{5}\gamma_{0}= & 0;\nonumber \\
\gamma_{l}\gamma_{5}+\gamma_{5}\gamma_{l}= & 0;\,\,\,\gamma_{5}^{2}=-1.\label{eq:32}
\end{align}
It shows that the matrix $\gamma_{5}$ is pseudo scalar matrix. Furthermore,
the quaternionic Dirac spinor $\psi=\psi_{0}+e_{1}\psi_{1}+e_{2}\psi_{2}+e_{3}\psi_{3}$
can now be decomposed as 

\begin{align}
\psi=\left(\begin{array}{c}
\psi_{a}\\
\psi_{b}
\end{array}\right)= & \left(\begin{array}{c}
\psi_{0}\\
\psi_{1}\\
\psi_{2}\\
-\psi_{3}
\end{array}\right)\label{eq:33}
\end{align}
in terms of two and four components Dirac spinors associated with
symplectic representation of quaternions $\psi=\psi_{a}+e_{2}\psi_{b}$with
$\psi_{a}=\psi_{0}+e_{1}\psi_{1}$ and $\psi_{b}=\psi_{2}-e_{1}\psi_{3}$.
Furthermore, we may also write one component quaternion valued Dirac
spinor which is isomorphic to two component complex spinor and four
component real spinor representation. Substituting the value of $D_{\mu}$
from equation (\ref{eq:17}\textbf{\uline{47}}), we get 

\begin{align}
i\gamma_{\mu}\left(\partial_{\mu}\psi\left(x,t\right)+\mathbf{e}A_{\mu}\psi\left(x,t\right)-\mathbf{g}\psi\left(x,t\right)B_{\mu}\right)= & m\psi\left(x,t\right).\label{eq:34}
\end{align}
Splitting $\gamma_{\mu}$ ,$\partial_{\mu}$,$A_{\mu}$and $B_{\mu}$
in terms of real and quaternionic constituents, we get

\begin{align}
i\gamma_{0}\left(\partial_{0}\psi+\mathbf{e}A_{0}\psi-\mathbf{g}\psi B_{0}\right)+i\gamma_{l}\left(\partial_{l}\psi+\mathbf{e}A_{l}\psi-\mathbf{g}\psi B_{l}\right)= & m\psi;\label{eq:35}
\end{align}
which is the general equation of spin-$\frac{1}{2}$ particle (dyon)
in generalized electromagnetic field. Equation (\ref{eq:35}) may
now be reduced as 

\begin{align}
i\gamma_{0}\left(-iE\psi+\mathbf{e}A_{0}\psi-\mathbf{g}\psi B_{0}\right)+i\gamma_{l}\left(ip_{l}\psi+\mathbf{e\,}e_{l}A_{l}\psi-\mathbf{g}\psi e_{l}B_{l}\right)= & m\psi\label{eq:36}
\end{align}
which can also be written explicitly as

\begin{align}
\left[\begin{array}{cc}
1 & 0\\
0 & -1
\end{array}\right]\left(E\psi+i\mathbf{e}A_{0}\psi-i\mathbf{g}\psi B_{0}\right) & +\nonumber \\
\left[\begin{array}{cc}
0 & ie_{l}\\
-ie_{l} & 0
\end{array}\right]\left(-P_{l}\psi+i\mathbf{e}e_{l}A_{l}\psi-i\mathbf{g}\psi e_{l}B_{l}\right)-m\psi & =0\label{eq:37}
\end{align}
Let us study the above equation for different cases

\subsection{Case (a) For electric field due to electric charge}

Let us discuss the case when we have only pure electric field associated
with electric charge $\mathbf{e}$. In this case we have $A_{0}\neq0\,,\, A_{l}=0,\, B_{\mu}=0$
so that the equation (\ref{eq:37}) reduces to

\begin{alignat}{1}
\left[\begin{array}{cc}
1 & 0\\
0 & -1
\end{array}\right]\left(E+i\,\mathbf{e}\, A_{0}\right)\left(\begin{array}{c}
\psi_{a}\\
\psi_{b}
\end{array}\right)+\left[\begin{array}{cc}
0 & ie_{l}.P_{l}\\
-ie_{l}.p_{l} & 0
\end{array}\right]\left(\begin{array}{c}
\psi_{a}\\
\psi_{b}
\end{array}\right)-m\left(\begin{array}{c}
\psi_{a}\\
\psi_{b}
\end{array}\right)= & 0\label{eq:38}
\end{alignat}
which further reproduces two coupled equations

\begin{align}
\mathcal{\widehat{A}^{\dagger}}\psi_{b}=ie_{l}.P_{l}\psi_{b}=\left(E+i\,\mathbf{e\,}A_{0}-m\right) & \psi_{a};\nonumber \\
\mathcal{\widehat{A}}\psi_{a}=ie_{l}.P_{l}\psi_{a}=\left(E+i\,\mathbf{e}\, A_{0}+m\right) & \psi_{b};\label{eq:39}
\end{align}
where $\mathcal{\widehat{A}}=\mathcal{\widehat{A}^{\dagger}}=ie_{l}p_{l}$.These
two-coupled equations (\ref{eq:39}) can now be decoupled into a single
equation leading to its supersymmetrization as 

\begin{align}
P_{l}^{2}\psi_{a,b}= & \left\{ \left(E+i\,\mathbf{e\,}A_{0}\right)^{2}-m^{2}\right\} \psi_{a,b}\label{eq:40}
\end{align}
so that the super partner Hamiltonian may now be written as

\begin{align}
\mathcal{\widehat{H}}_{-}= & \mathcal{\widehat{A}^{\dagger}}\mathcal{\widehat{A}}=P_{l}^{2};\nonumber \\
\mathcal{\widehat{H}}_{+}= & \mathcal{\widehat{A}}\mathcal{\widehat{A}^{\dagger}}=P_{l}^{2}.\label{eq:41}
\end{align}
Corresponding Dirac Hamiltonian may be defined in the following manner
where we have used the Pauli-Dirac representation i.e. 

\begin{align}
\mathcal{\widehat{H}}_{D}= & \left[\begin{array}{cc}
m & ie_{l}P_{l}\\
ie_{l}P_{l} & -m
\end{array}\right].\label{eq:42}
\end{align}
Let us write equation (\ref{eq:42}) as compared to the standard Dirac
Hamiltonian given by Thaller \cite{key-21} as

\begin{alignat}{1}
\mathcal{\widehat{H}}_{D}= & \left[\begin{array}{cc}
M_{+} & \hat{Q}_{D}^{\dagger}\\
\hat{Q}_{D} & M_{-}
\end{array}\right]\label{eq:43}
\end{alignat}
which leads to $M_{+}=M_{-}=0$ and $\hat{Q}_{D}=\hat{Q}_{D}^{+}=ie_{l}P_{l}$
along with the following supersymmetric conditions

\begin{align}
\hat{Q}_{D}^{\dagger}M_{-}= & M_{+}\hat{Q}_{D}^{\dagger};\nonumber \\
\hat{Q}_{D}M_{+}= & M_{-}\hat{Q}_{D}\label{eq:44}
\end{align}
and the following expression for the square of the Dirac Hamiltonian
i.e. 

\begin{align}
\mathcal{\widehat{H}}_{D}^{2}= & \left[\begin{array}{cc}
\left(P_{l}^{2}+m^{2}\right) & 0\\
0 & \left(P_{l}^{2}+m^{2}\right)
\end{array}\right].\label{eq:45}
\end{align}
As such, we may write the Schrodinger Hamiltonian $\hat{H}_{s}$ and
Supercharges $\hat{Q}_{s}$ and $\hat{Q}_{s}^{\dagger}$ as

\begin{align}
\hat{H}_{s}= & \left[\begin{array}{cc}
P_{l}^{2} & 0\\
0 & P_{l}^{2}
\end{array}\right];\nonumber \\
\hat{Q}_{s}= & \left[\begin{array}{cc}
0 & ie_{l}P_{l}\\
0 & 0
\end{array}\right];\nonumber \\
\hat{Q}_{s}^{\dagger}= & \left[\begin{array}{cc}
0 & 0\\
ie_{l}P_{l} & 0
\end{array}\right];\label{eq:46}
\end{align}
which satisfy the following well known forms of supersymmetric (SUSY)
algebra i.e.

\begin{align}
\left[\hat{Q}_{s},\hat{H}_{s}\right]=\left[\hat{Q}_{s}^{\dagger},\hat{H}_{s}\right]= & 0\nonumber \\
\left\{ \hat{Q}_{s},\hat{Q}_{s}\right\} =\left\{ \hat{Q}_{s}^{\dagger},\hat{Q}_{s}^{\dagger}\right\} = & 0\nonumber \\
\left[\hat{Q}_{s},\hat{Q}_{s}^{\dagger}\right]= & \hat{H}_{s}^{+}.\label{eq:47}
\end{align}
We may also obtain the following types of four spinor amplitudes of
Dirac spinors i.e.
\begin{itemize}
\item One component spinor amplitudes
\end{itemize}
\begin{eqnarray}
\Psi^{1}= & (1+e_{2}.\frac{ie_{l}P_{l}}{E_{+}-\mathbf{e}\, A_{0}+m}) & \,(Energy=+ive,\, spin=\uparrow);\nonumber \\
\Psi^{2}= & (1+e_{2}.\frac{ie_{l}P_{l}}{E_{+}-\mathbf{e}\, A_{0}+m})e_{1} & \,(Energy=+ive,\, spin=\downarrow);\nonumber \\
\Psi^{3}= & (e_{2}-\frac{ie_{l}P_{l}}{E_{-}+\mathbf{e}\, A_{0}+m}) & \,(Energy=-ive,\, spin=\uparrow);\nonumber \\
\Psi^{4}= & (e_{2}-\frac{ie_{l}P_{l}}{E_{-}+\mathbf{e}\, A_{0}+m})e_{1} & \,(Energy=-ive,\, spin=\downarrow).\label{eq:48}
\end{eqnarray}

\begin{itemize}
\item Two component spinor amplitudes
\end{itemize}
\begin{eqnarray}
\Psi^{1} & = & \left(\begin{array}{c}
1\\
\frac{ie_{l}P_{l}}{E_{+}-\mathbf{e}\, A_{0}+m}
\end{array}\right)\,(Energy=+ive,\, spin=\uparrow);\nonumber \\
\Psi^{2} & = & \left(\begin{array}{c}
1\\
\frac{ie_{l}P_{l}}{E_{+}-\mathbf{e}\, A_{0}+m}
\end{array}\right)e_{1}\,(Energy=+ive,\, spin=\downarrow);\nonumber \\
\Psi^{3} & = & \left(\begin{array}{c}
-\frac{ie_{l}P_{l}}{E_{-}+\mathbf{e}\, A_{0}+m}\\
1
\end{array}\right)\,(Energy=-ive,\, spin=\uparrow);\nonumber \\
\Psi^{4} & = & \left(\begin{array}{c}
-\frac{ie_{l}P_{l}}{E_{-}+\mathbf{e}\, A_{0}+m}\\
1
\end{array}\right)e_{1}\,(Energy=-ive,\, spin=\downarrow).\label{eq:49}
\end{eqnarray}

\begin{itemize}
\item Four component spinor amplitudes may also be obtained by restricting
the direction of propagation along any one axis which we suppose $Z-axis$
i.e ($p_{x}=p_{y}=0)$ and on substituting $e_{l}=-i\sigma_{l}$ and
$\sigma_{1}=\left(\begin{array}{cc}
0 & 1\\
1 & 0
\end{array}\right),\,\,\sigma_{2}=\left(\begin{array}{cc}
0 & -i\\
i & 0
\end{array}\right)\,\,\sigma_{3}=\left(\begin{array}{cc}
1 & 0\\
0 & -1
\end{array}\right)$ along with the usual definitions of spin up and spin down amplitudes
of spin i.e.
\end{itemize}
\begin{eqnarray}
\psi^{1} & = & \left(\begin{array}{c}
1\\
0\\
\frac{\left|\vec{p}\right|}{E_{+}-\mathbf{e}\, A_{0}+m}\\
0
\end{array}\right)(Energy=+ive,\, spin=\uparrow);\nonumber \\
\psi^{2} & = & \left(\begin{array}{c}
0\\
1\\
0\\
-\frac{\left|\vec{p}\right|}{E_{+}-\mathbf{e}\, A_{0}+m}
\end{array}\right)(Energy=+ive,\, spin=\downarrow);\nonumber \\
\psi^{3} & = & \left(\begin{array}{c}
-\frac{\left|\vec{p}\right|}{E_{-}+\mathbf{e}\, A_{0}+m}\\
0\\
1\\
0
\end{array}\right)(Energy=-ive,\, spin=\uparrow);\nonumber \\
\psi^{4} & = & \left(\begin{array}{c}
0\\
\frac{\left|\vec{p}\right|}{E_{-}+\mathbf{e}\, A_{0}+m}\\
0\\
1
\end{array}\right)(Energy=-ive,\, spin=\downarrow).\label{eq:50}
\end{eqnarray}
As such, we have obtained the solution of quaternion Dirac equation
for dyons in terms of one component quaternion, two component complex
and four component real spinor amplitudes. Equation (\ref{eq:50})
is same as obtained for the case of usual Dirac equation in electromagnetic
field. Thus we may interpret that the $N=1$ quaternion spinor amplitude
is isomorphic to $N=2$ complex and $N=4$ real spinor amplitude solution
of Dirac equation for dyons. We can accordingly interpret the minimum
dimensional representation for Dirac equation is $N=1$ in quaternionic
case, $N=2$ in complex case and $N=4$ for real number field.

\subsection{Case (b): For magnetic field due to electric charge}

Let us discuss the case when we have only pure magnetic associated
with electric charge $\mathbf{e}$. In this case we have $A_{0}=0\,,\, A_{l}\neq0,\, B_{\mu}=0$
so that the equation (\ref{eq:37}) reduces to

\begin{align}
\left[\begin{array}{cc}
1 & 0\\
0 & -1
\end{array}\right]E\left(\begin{array}{c}
\psi_{a}\\
\psi_{b}
\end{array}\right)+\left[\begin{array}{cc}
0 & ie_{l}\\
-ie_{l} & 0
\end{array}\right](-P_{l}+i\,\mathbf{e}e_{l}A_{l})\left(\begin{array}{c}
\psi_{a}\\
\psi_{b}
\end{array}\right)-m\left(\begin{array}{c}
\psi_{a}\\
\psi_{b}
\end{array}\right)= & 0\label{eq:51}
\end{align}
which yields two coupled equations i.e.

\begin{align}
\mathcal{\widehat{A}^{\dagger}}\psi_{b}=ie_{l}(P_{l}-i\,\mathbf{e}e_{l}A_{l})\psi_{b}=\left(E-m\right) & \psi_{a};\nonumber \\
\mathcal{\widehat{A}}\psi_{a}=ie_{l}(P_{l}-i\,\mathbf{e}e_{l}A_{l})\psi_{a}=\left(E+m\right) & \psi_{b};\label{eq:52}
\end{align}
where $\mathcal{\widehat{A}}=\mathcal{\widehat{A}^{\dagger}}=ie_{l}(P_{l}-i\,\mathbf{\,}e_{l}A_{l}).$
These two-coupled equations can be decoupled into a single coupled
equation showing supersymmetry in the following manner 
\begin{align}
[ie_{l}(P_{l}-i\,\mathbf{e}e_{l}A_{l})]^{2}\psi_{a,b}= & \left\{ E-m^{2}\right\} \psi_{a,b}\label{eq:53}
\end{align}
so that the super partner Hamiltonian may now be written as

\begin{align}
\mathcal{\widehat{H}}_{-}= & \mathcal{\widehat{A}^{\dagger}}\mathcal{\widehat{A}}=\mathcal{\widehat{H}}_{+}=\mathcal{\widehat{A}}\mathcal{\widehat{A}^{\dagger}}=[ie_{l}(P_{l}-i\,\mathbf{e}e_{l}A_{l})]^{2}.\label{eq:54}
\end{align}
Thus the corresponding Dirac Hamiltonian may be defined in the following
manner 

\begin{align}
\mathcal{\widehat{H}}_{D}= & \left[\begin{array}{cc}
m & ie_{l}(P_{l}-i\,\mathbf{e}e_{l}A_{l})\\
ie_{l}(P_{l}-i\,\mathbf{e}e_{l}A_{l}) & -m
\end{array}\right].\label{eq:55}
\end{align}
Like wise, the previous case of electric field, here in case of magnetic
field we may also obtain $M_{+}=M_{-}=m$ and $\hat{Q}_{D}=\hat{Q}_{D}^{+}=ie_{l}(P_{l}-i\,\mathbf{e}e_{l}A_{l})$
along with the supersymmetric condition (\ref{eq:44}) and the following
expression for the square of the Dirac Hamiltonian as 

\begin{align}
\mathcal{\widehat{H}}_{D}^{2}= & \left[\begin{array}{cc}
[ie_{l}(P_{l}-i\,\mathbf{e}e_{l}A_{l})]^{2}+m^{2} & 0\\
0 & [ie_{l}(P_{l}-i\,\mathbf{e}e_{l}A_{l})]^{2}+m^{2}
\end{array}\right].\label{eq:56}
\end{align}
Accordingly we may write the Schrodinger Hamiltonian $\hat{H}_{s}$
and Supercharges $\hat{Q}_{s}$ and $\hat{Q}_{s}^{\dagger}$ as

\begin{align}
\hat{H}_{s}= & \left[\begin{array}{cc}
[ie_{l}(P_{l}-i\,\mathbf{e}e_{l}A_{l})]^{2} & 0\\
0 & [ie_{l}(P_{l}-i\,\mathbf{e}e_{l}A_{l})]^{2}
\end{array}\right];\nonumber \\
\hat{Q}_{s}= & \left[\begin{array}{cc}
0 & [ie_{l}(P_{l}-i\,\mathbf{e}e_{l}A_{l})]\\
0 & 0
\end{array}\right];\nonumber \\
\hat{Q}_{s}^{\dagger}= & \left[\begin{array}{cc}
0 & 0\\
{}[ie_{l}(P_{l}-i\,\mathbf{e}e_{l}A_{l})] & 0
\end{array}\right].\label{eq:57}
\end{align}
Here, also $\hat{H}_{s}$, $\hat{Q}_{s}$ and $\hat{Q}_{s}^{\dagger}$
satisfy the well known supersymmetric (SUSY) algebra given by equation
(\ref{eq:47}). Consequently, we may also obtain the following types
of four spinor amplitudes of Dirac spinors in presence of pure magnetic
field as i.e.
\begin{itemize}
\item One component spinor amplitudes
\end{itemize}
\begin{eqnarray}
\Psi^{1}= & (1+e_{2}.\frac{[ie_{l}(P_{l}-i\,\mathbf{e}e_{l}A_{l})]}{E_{+}+m}) & \,(Energy=+ive,\, spin=\uparrow);\nonumber \\
\Psi^{2}= & (1+e_{2}.\frac{[ie_{l}(P_{l}-i\,\mathbf{e}e_{l}A_{l})]}{E_{+}+m})e_{1} & \,(Energy=+ive,\, spin=\downarrow);\nonumber \\
\Psi^{3}= & (e_{2}-\frac{[ie_{l}(P_{l}-i\,\mathbf{e}e_{l}A_{l})]}{E_{-}+m}) & \,(Energy=-ive,\, spin=\uparrow);\nonumber \\
\Psi^{4}= & (e_{2}-\frac{[ie_{l}(P_{l}-i\,\mathbf{e}e_{l}A_{l})]}{E_{-}+m})e_{1} & \,(Energy=-ive,\, spin=\downarrow).\label{eq:58}
\end{eqnarray}

\begin{itemize}
\item Two component spinor amplitudes
\end{itemize}
\begin{eqnarray}
\Psi^{1} & = & \left(\begin{array}{c}
1\\
\frac{[ie_{l}(P_{l}-i\,\mathbf{e}e_{l}A_{l})]}{E_{+}+m}
\end{array}\right)\,(Energy=+ive,\, spin=\uparrow);\nonumber \\
\Psi^{2} & = & \left(\begin{array}{c}
1\\
\frac{[ie_{l}(P_{l}-i\,\mathbf{e}e_{l}A_{l})]}{E_{+}+m}
\end{array}\right)e_{1}\,(Energy=+ive,\, spin=\downarrow);\nonumber \\
\Psi^{3} & = & \left(\begin{array}{c}
-\frac{[ie_{l}(P_{l}-i\,\mathbf{e}e_{l}A_{l})]}{E_{-}+m}\\
1
\end{array}\right)\,(Energy=-ive,\, spin=\uparrow);\nonumber \\
\Psi^{4} & = & \left(\begin{array}{c}
-\frac{[ie_{l}(P_{l}-i\,\mathbf{e}e_{l}A_{l})]}{E_{-}+m}\\
1
\end{array}\right)e_{1}\,(Energy=-ive,\, spin=\downarrow).\label{eq:59}
\end{eqnarray}

\begin{itemize}
\item Four component spinor amplitudes may also be obtained by restricting
the direction of propagation along any one axis which we suppose $Z-axis$
i.e ($p_{x}=p_{y}=0)$ and $(A_{x}=A_{y}=0\Rightarrow H_{z}=0)$.
Accordingly, substituting $e_{l}=-i\sigma_{l}$ and $\sigma_{1}=\left(\begin{array}{cc}
0 & 1\\
1 & 0
\end{array}\right),\,\,\sigma_{2}=\left(\begin{array}{cc}
0 & -i\\
i & 0
\end{array}\right)\,\,\sigma_{3}=\left(\begin{array}{cc}
1 & 0\\
0 & -1
\end{array}\right)$ along with the usual definitions of spin up and spin down amplitudes
of spin , we get
\end{itemize}
\begin{eqnarray}
\psi^{1} & = & \left(\begin{array}{c}
1\\
0\\
\frac{\left|\vec{p}\right|}{E_{+}+m}\\
0
\end{array}\right)(Energy=+ive,\, spin=\uparrow);\nonumber \\
\psi^{2} & = & \left(\begin{array}{c}
0\\
1\\
0\\
-\frac{\left|\vec{p}\right|}{E_{+}+m}
\end{array}\right)(Energy=+ive,\, spin=\downarrow);\nonumber \\
\psi^{3} & = & \left(\begin{array}{c}
-\frac{\left|\vec{p}\right|}{E_{-}+m}\\
0\\
1\\
0
\end{array}\right)(Energy=-ive,\, spin=\uparrow);\nonumber \\
\psi^{4} & = & \left(\begin{array}{c}
0\\
\frac{\left|\vec{p}\right|}{E_{-}+m}\\
0\\
1
\end{array}\right)(Energy=-ive,\, spin=\downarrow).\label{eq:60}
\end{eqnarray}
which are the well known usual spinor amplitudes for a Dirac free
Particle .

\subsection{Case (c): For Electric field due to magnetic monopole}

Here, we discuss the case when we have only electric field associated
with magnetic charge (pure magnetic monopole) $\mathbf{g}$ only.
So, by virtue of duality of magnetic charge \cite{key-27,key-28,key-29,key-30},
we take $B_{0}=0\,,\, B_{l}\neq0,\, A_{\mu}=0$. Thus, the equation
(\ref{eq:37}) reduces to

\begin{align}
\left[\begin{array}{cc}
1 & 0\\
0 & -1
\end{array}\right]E\left(\begin{array}{c}
\psi_{a}\\
\psi_{b}
\end{array}\right)+\left[\begin{array}{cc}
0 & ie_{l}\\
-ie_{l} & 0
\end{array}\right](-P_{l}+i\,\mathbf{g\,}e_{l}B_{l})\left(\begin{array}{c}
\psi_{a}\\
\psi_{b}
\end{array}\right)-m\left(\begin{array}{c}
\psi_{a}\\
\psi_{b}
\end{array}\right)= & 0\label{eq:61}
\end{align}
which yields two coupled equations i.e.

\begin{align}
\mathcal{\widehat{A}^{\dagger}}\psi_{b}=ie_{l}(P_{l}-i\,\mathbf{g\,}e_{l}B_{l})\psi_{b}=\left(E-m\right) & \psi_{a};\nonumber \\
\mathcal{\widehat{A}}\psi_{a}=ie_{l}(P_{l}-i\,\mathbf{g}e_{l}B_{l})\psi_{a}=\left(E+m\right) & \psi_{b};\label{eq:62}
\end{align}
where $\mathcal{\widehat{A}}=\mathcal{\widehat{A}^{\dagger}}=ie_{l}(P_{l}-i\,\mathbf{g\,}e_{l}B_{l}).$
These two-coupled equations can be decoupled into a single coupled
equation showing supersymmetry in the following manner 
\begin{align}
[ie_{l}(P_{l}-i\,\mathbf{g\,}e_{l}B_{l})]^{2}\psi_{a,b}= & \left\{ E-m^{2}\right\} \psi_{a,b}\label{eq:63}
\end{align}
so that the super partner Hamiltonian may now be written as

\begin{align}
\mathcal{\widehat{H}}_{-}= & \mathcal{\widehat{A}^{\dagger}}\mathcal{\widehat{A}}=\mathcal{\widehat{H}}_{+}=\mathcal{\widehat{A}}\mathcal{\widehat{A}^{\dagger}}=[ie_{l}(P_{l}-i\,\mathbf{g\,}e_{l}B_{l})]^{2}.\label{eq:64}
\end{align}
Thus the corresponding Dirac Hamiltonian may be defined in the following
manner 

\begin{align}
\mathcal{\widehat{H}}_{D}= & \left[\begin{array}{cc}
m & ie_{l}(P_{l}-i\,\mathbf{g\,}e_{l}B_{l})\\
ie_{l}(P_{l}-i\,\mathbf{g\,}e_{l}B_{l}) & -m
\end{array}\right].\label{eq:65}
\end{align}
Like wise, the previous case of electric field, here in case of magnetic
field we may also obtain $M_{+}=M_{-}=m$ and $\hat{Q}_{D}=\hat{Q}_{D}^{+}=ie_{l}(P_{l}-i\,\mathbf{g\,}e_{l}B_{l})$
along with the supersymmetric condition (\ref{eq:44}) and the following
expression for the square of the Dirac Hamiltonian as 

\begin{align}
\mathcal{\widehat{H}}_{D}^{2}= & \left[\begin{array}{cc}
[ie_{l}(P_{l}-i\,\mathbf{g\,}e_{l}B_{l})]^{2}+m^{2} & 0\\
0 & [ie_{l}(P_{l}-i\,\mathbf{g\,}e_{l}B_{l})+m^{2}
\end{array}\right].\label{eq:66}
\end{align}
Accordingly we may write the Schrodinger Hamiltonian $\hat{H}_{s}$
and Supercharges $\hat{Q}_{s}$ and $\hat{Q}_{s}^{\dagger}$ as

\begin{align}
\hat{H}_{s}= & \left[\begin{array}{cc}
[ie_{l}(P_{l}-i\,\mathbf{g\,}e_{l}B_{l})]^{2} & 0\\
0 & [ie_{l}(P_{l}-i\,\mathbf{g\,}e_{l}B_{l})]^{2}
\end{array}\right];\nonumber \\
\hat{Q}_{s}= & \left[\begin{array}{cc}
0 & [ie_{l}(P_{l}-i\,\mathbf{g\,}e_{l}B_{l})]\\
0 & 0
\end{array}\right];\nonumber \\
\hat{Q}_{s}^{\dagger}= & \left[\begin{array}{cc}
0 & 0\\
{}[ie_{l}(P_{l}-i\,\mathbf{g\,}e_{l}B_{l})] & 0
\end{array}\right].\label{eq:67}
\end{align}
Here, also $\hat{H}_{s}$, $\hat{Q}_{s}$ and $\hat{Q}_{s}^{\dagger}$
satisfy the well known supersymmetric (SUSY) algebra given by equation
(\ref{eq:47}). Consequently, we may also obtain the following types
of four spinor amplitudes of Dirac spinors in presence of pure magnetic
field as i.e.
\begin{itemize}
\item One component spinor amplitudes
\end{itemize}
\begin{eqnarray}
\Psi^{1}= & (1+e_{2}.\frac{[ie_{l}(P_{l}-i\,\mathbf{g\,}e_{l}B_{l})]}{E_{+}+m}) & \,(Energy=+ive,\, spin=\uparrow);\nonumber \\
\Psi^{2}= & (1+e_{2}.\frac{[ie_{l}(P_{l}-i\,\mathbf{g\,}e_{l}B_{l})]}{E_{+}+m})e_{1} & \,(Energy=+ive,\, spin=\downarrow);\nonumber \\
\Psi^{3}= & (e_{2}-\frac{[ie_{l}(P_{l}-i\,\mathbf{g\,}e_{l}B_{l})]}{E_{-}+m}) & \,(Energy=-ive,\, spin=\uparrow);\nonumber \\
\Psi^{4}= & (e_{2}-\frac{[ie_{l}(P_{l}-i\,\mathbf{g\,}e_{l}B_{l})]}{E_{-}+m})e_{1} & \,(Energy=-ive,\, spin=\downarrow).\label{eq:68}
\end{eqnarray}

\begin{itemize}
\item Two component spinor amplitudes
\end{itemize}
\begin{eqnarray}
\Psi^{1} & = & \left(\begin{array}{c}
1\\
\frac{[ie_{l}(P_{l}-i\,\mathbf{g\,}e_{l}B_{l})]}{E_{+}+m}
\end{array}\right)\,(Energy=+ive,\, spin=\uparrow);\nonumber \\
\Psi^{2} & = & \left(\begin{array}{c}
1\\
\frac{[ie_{l}(P_{l}-i\,\mathbf{g\,}e_{l}B_{l}))]}{E_{+}+m}
\end{array}\right)e_{1}\,(Energy=+ive,\, spin=\downarrow);\nonumber \\
\Psi^{3} & = & \left(\begin{array}{c}
-\frac{[ie_{l}(P_{l}-i\,\mathbf{e}e_{l}A_{l})]}{E_{-}+m}\\
1
\end{array}\right)\,(Energy=-ive,\, spin=\uparrow);\nonumber \\
\Psi^{4} & = & \left(\begin{array}{c}
-\frac{[ie_{l}(P_{l}-i\,\mathbf{g\,}e_{l}B_{l}))]}{E_{-}+m}\\
1
\end{array}\right)e_{1}\,(Energy=-ive,\, spin=\downarrow).\label{eq:69}
\end{eqnarray}

\begin{itemize}
\item Four component spinor amplitudes may also be obtained by restricting
the direction of propagation along any one axis which we suppose $Z-axis$
i.e ($p_{x}=p_{y}=0)$ and $(B_{x}=B_{y}=0\Rightarrow E_{z}=0)$.
Accordingly, substituting $e_{l}=-i\sigma_{l}$ and $\sigma_{1}=\left(\begin{array}{cc}
0 & 1\\
1 & 0
\end{array}\right),\,\,\sigma_{2}=\left(\begin{array}{cc}
0 & -i\\
i & 0
\end{array}\right)\,\,\sigma_{3}=\left(\begin{array}{cc}
1 & 0\\
0 & -1
\end{array}\right)$ along with the usual definitions of spin up and spin down amplitudes
of spin , we get the four component Dirac spinors same as the equation
(\ref{eq:60}).
\end{itemize}

\subsection{Case (d): For Magnetic field due to magnetic monopole}

Let us discuss the case when we have only pure magnetic field associated
with magnetic charge (monopole) $\mathbf{g}$. In this case we have
$B_{0}\neq0\,,\, B_{l}=0,\, A_{\mu}=0$ so that the equation (\ref{eq:37})
reduces to

\begin{alignat}{1}
\left[\begin{array}{cc}
1 & 0\\
0 & -1
\end{array}\right]\left(E+i\,\mathbf{g}\, B_{0}\right)\left(\begin{array}{c}
\psi_{a}\\
\psi_{b}
\end{array}\right)+\left[\begin{array}{cc}
0 & ie_{l}.P_{l}\\
-ie_{l}.p_{l} & 0
\end{array}\right]\left(\begin{array}{c}
\psi_{a}\\
\psi_{b}
\end{array}\right)-m\left(\begin{array}{c}
\psi_{a}\\
\psi_{b}
\end{array}\right)= & 0\label{eq:70}
\end{alignat}
which further reduces to two coupled equations

\begin{align}
\mathcal{\widehat{A}^{\dagger}}\psi_{b}=ie_{l}.P_{l}\psi_{b}=\left(E+i\,\mathbf{g\,}B_{0}-m\right) & \psi_{a};\nonumber \\
\mathcal{\widehat{A}}\psi_{a}=ie_{l}.P_{l}\psi_{a}=\left(E+i\,\mathbf{g\,}B_{0}-m\right) & \psi_{b};\label{eq:71}
\end{align}
where $\mathcal{\widehat{A}}=\mathcal{\widehat{A}^{\dagger}}=ie_{l}p_{l}$.These
two-coupled equations (\ref{eq:71}) can now be decoupled into a single
equation leading to its supersymmetrization as 

\begin{align}
P_{l}^{2}\psi_{a,b}= & \left\{ \left(E+i\,\mathbf{g\,}B_{0}\right)^{2}-m^{2}\right\} \psi_{a,b};\label{eq:72}
\end{align}
so that the super partner Hamiltonian may now be written as equation
(\ref{eq:41}). Corresponding Dirac Hamiltonian then may be defined
as equation (\ref{eq:42}) which can also be written as (\ref{eq:43})
after its comparison with the standard Dirac Hamiltonian given by
Thaller \cite{key-21} and thus, leads to $M_{+}=M_{-}=0$ and $\hat{Q}_{D}=\hat{Q}_{D}^{+}=ie_{l}P_{l}$
along with the following supersymmetric conditions given by equation
(\ref{eq:44}) along with the Dirac Hamiltonian given by 

\begin{align}
\mathcal{\widehat{H}}_{D}^{2}= & \left[\begin{array}{cc}
\left(P_{l}^{2}+m^{2}\right) & 0\\
0 & \left(P_{l}^{2}+m^{2}\right)
\end{array}\right]=\widehat{H}_{s}^{2}+m^{2}\widehat{I}\label{eq:73}
\end{align}
where $\hat{I}$ is unit matrix of order $4$. Consequently, we may
write the Schrodinger Hamiltonian $\hat{H}_{s}$ and Supercharges
( $\hat{Q}_{s}$ and $\hat{Q}_{s}^{\dagger}$) as given by equation
(\ref{eq:46} ) leading to well known supersymmetric (SUSY) algebra
relations given by equation (\ref{eq:47}). Furthermore, the following
types of four spinor amplitudes of Dirac spinors may also be obtained
as 
\begin{itemize}
\item One component spinor amplitudes
\end{itemize}
\begin{eqnarray}
\Psi^{1}= & (1+e_{2}.\frac{ie_{l}P_{l}}{\left(E_{+}-\mathbf{g\,}B_{0}+m\right)}) & \,(Energy=+ive,\, spin=\uparrow);\nonumber \\
\Psi^{2}= & (1+e_{2}.\frac{ie_{l}P_{l}}{\left(E_{+}-\mathbf{g\,}B_{0}+m\right)})e_{1} & \,(Energy=+ive,\, spin=\downarrow);\nonumber \\
\Psi^{3}= & (e_{2}-\frac{ie_{l}P_{l}}{\left(E_{-}+\mathbf{g\,}B_{0}+m\right)}) & \,(Energy=-ive,\, spin=\uparrow);\nonumber \\
\Psi^{4}= & (e_{2}-\frac{ie_{l}P_{l}}{\left(E_{-}+i\,\mathbf{g\,}B_{0}+m\right)})e_{1} & \,(Energy=-ive,\, spin=\downarrow).\label{eq:74}
\end{eqnarray}

\begin{itemize}
\item Two component spinor amplitudes
\end{itemize}
\begin{eqnarray}
\Psi^{1} & = & \left(\begin{array}{c}
1\\
\frac{ie_{l}P_{l}}{\left(E_{+}-\mathbf{g\,}B_{0}+m\right)}
\end{array}\right)\,(Energy=+ive,\, spin=\uparrow);\nonumber \\
\Psi^{2} & = & \left(\begin{array}{c}
1\\
\frac{ie_{l}P_{l}}{\left(E_{+}-\mathbf{g\,}B_{0}+m\right)}
\end{array}\right)e_{1}\,(Energy=+ive,\, spin=\downarrow);\nonumber \\
\Psi^{3} & = & \left(\begin{array}{c}
-\frac{ie_{l}P_{l}}{\left(E_{-}+\mathbf{g\,}B_{0}+m\right)}\\
1
\end{array}\right)\,(Energy=-ive,\, spin=\uparrow);\nonumber \\
\Psi^{4} & = & \left(\begin{array}{c}
-\frac{ie_{l}P_{l}}{\left(E_{-}+\mathbf{g\,}B_{0}+m\right)}\\
1
\end{array}\right)e_{1}\,(Energy=-ive,\, spin=\downarrow).\label{eq:75}
\end{eqnarray}

\begin{itemize}
\item Four component spinor amplitudes may also be obtained by restricting
the direction of propagation along any one axis which we suppose $Z-axis$
i.e ($p_{x}=p_{y}=0)$ and on substituting $e_{l}=-i\sigma_{l}$ and
$\sigma_{1}=\left(\begin{array}{cc}
0 & 1\\
1 & 0
\end{array}\right),\,\,\sigma_{2}=\left(\begin{array}{cc}
0 & -i\\
i & 0
\end{array}\right)\,\,\sigma_{3}=\left(\begin{array}{cc}
1 & 0\\
0 & -1
\end{array}\right)$ along with the usual definitions of spin up and spin down amplitudes
of spin i.e.
\end{itemize}
\begin{eqnarray}
\psi^{1} & = & \left(\begin{array}{c}
1\\
0\\
\frac{\left|\vec{p}\right|}{\left(E_{+}-\mathbf{g\,}B_{0}+m\right)}\\
0
\end{array}\right)(Energy=+ive,\, spin=\uparrow);\nonumber \\
\psi^{2} & = & \left(\begin{array}{c}
0\\
1\\
0\\
-\frac{\left|\vec{p}\right|}{\left(E_{+}-\mathbf{g\,}B_{0}+m\right)}
\end{array}\right)(Energy=+ive,\, spin=\downarrow);\nonumber \\
\psi^{3} & = & \left(\begin{array}{c}
-\frac{\left|\vec{p}\right|}{\left(E_{-}+g\, B_{0}+m\right)}\\
0\\
1\\
0
\end{array}\right)(Energy=-ive,\, spin=\uparrow);\nonumber \\
\psi^{4} & = & \left(\begin{array}{c}
0\\
\frac{\left|\vec{p}\right|}{\left(E_{-}+\mathbf{g\,}B_{0}+m\right)}\\
0\\
1
\end{array}\right)(Energy=-ive,\, spin=\downarrow).\label{eq:76}
\end{eqnarray}

\section{Discussion and Conclusion}

We have discussed the quaternion Dirac equation in electromagnetic
field where the partial derivative has been replaced by the quaternion
covariant derivative in terms of two gauge potentials. These two gauge
potentials are identified as the gauge potentials associated with
a particle which contains the simultaneous existence of electric and
magnetic charge (monopole). Such type of particles are named as dyons.
Thereafter, we have established consistently the SUSY for different
cases of quaternion Dirac equation for dyons. \textbf{Case (a) }deals
with the study of supersymmetrization of Dirac equation when it interacts
with electric field produced by electric charge only. Thereby, we
have obtained a single decoupled equation, super partner Hamiltonians,
total (Schrodinger) Hamiltonians and Schrodinger supercharges consistently
followed by the Dirac Hamiltonian and Dirac supercharges. It is shown
that the supercharges and Hamiltonian satisfy the SUSY algebra performing
SUSY transformations. Moreover, in this case, we have also obtained
the solutions of the Dirac equation for one component, two components
and four component Dirac spinors with various energy and spins. \textbf{Case
(b)} is described for a dyon consisting electric charge but moves
in only magnetic field. Likewise, we have followed the same procedure
and obtained consistently the particle Hamiltonian and supercharges
to satisfy the SUSY algebra. Furthermore, we have obtained the consistently
one component, two components and four component Dirac spinors with
various energy and spins. Same procedure has also been extended for
\textbf{Case (c)} and \textbf{Case (d)} respectively associated with
the electric and magnetic fields due to the presence of magnetic monopole
in order to establish the consistent formulation of SUSY and Dirac
spinors of various energy and spins.\textbf{ }It is concluded that
the \textbf{Case (a)} and \textbf{Case (d)} and likewise, \textbf{Case
(b)} and \textbf{Case (c)} are dual invariant. These cases may also
be analyzed by applying the duality transformations between electric
and magnetic constituents of dyons. It may also be concluded that
minimal representation for quaternion Dirac equation is described
as $N=1$ quaternionic, $N=2$complex and $N=4$ real representation.
In fact, the one-component spinor amplitudes are isomorphic to two
component complex spinor amplitudes and four component real spinor
amplitudes. As such, the higher dimensional supersymmetric Dirac equation
in generalized electromagnetic fields of dyons may be tackled well
in terms of quaternions splitting into$N=1$ quaternionic, $N=2$complex
and $N=4$ real representations of Supersymmetric quantum mechanics. 

\textbf{ACKNOWLEDGMENT}: One of us (OPSN) acknowledges the financial
support for UNESCO-TWAS Associateship from Third World Academy of
Sciences, Trieste (Italy) and Chinese Academy of Sciences, Beijing.
He is also thankful to ProfessorYue-Liang Wu\textbf{, }Director ITP
for his hospitality and research facilities\textbf{ }at ITP and KITP.

\end{document}